\documentclass[
 reprint,
superscriptaddress,
showpacs,
 amsmath,amssymb,
 aps,
]{revtex4-1}

\usepackage{graphicx,epsf}
\usepackage{float}
\usepackage{bm}
\usepackage{xcolor}

\usepackage{ulem} 

\begin{document}

\title{Conductance oscillations of core-shell nanowires in transversal magnetic fields}

\author{Andrei Manolescu}
\email{manoles@ru.is}
\affiliation{School of Science and Engineering, Reykjavik University, 
Menntavegur 1, IS-101 Reykjavik, Iceland}
\author{George Alexandru Nemnes}
\affiliation{University of Bucharest, Faculty of Physics, MDEO 
Research Center, 077125 Magurele-Ilfov, Romania}
\affiliation{Horia Hulubei National Institute for Physics and 
Nuclear Engineering, 077126 Magurele-Ilfov, Romania}
\author{Anna Sitek}
\affiliation{Science Institute, University of Iceland, Dunhaga 3, 
IS-107 Reykjavik, Iceland}
\affiliation{Department of Theoretical Physics, Faculty of Fundamental Problems of Technology,
Wroclaw University of Technology, 
50-370 Wroclaw, Poland} 
\author{Tomas~Orn Rosdahl}
\affiliation{Kavli Institute of Nanoscience, Delft University of Technology, 
2600 GA Delft, The Netherlands} 
\author{Sigurdur Ingi Erlingsson}
\affiliation{School of Science and Engineering, Reykjavik University, 
Menntavegur 1, IS-101 Reykjavik, Iceland}
\author{Vidar Gudmundsson}
\affiliation{Science Institute, University of Iceland, Dunhaga 3, 
              IS-107 Reykjavik, Iceland}


\begin{abstract} 
We analyze theoretically electronic transport through a core-shell nanowire in
the presence of a transversal magnetic field. We calculate the conductance
for a variable coupling between the nanowire and the attached
leads and show how the snaking states, which are low-energy states
localized along the lines of vanishing radial component of the magnetic
field, manifest their existence. In the strong coupling regime they
induce flux periodic, Aharonov-Bohm-like, conductance oscillations, which, by decreasing
the coupling to the leads, evolve into well resolved peaks. 
The flux periodic oscillations arise due to interference of the snaking states, 
which is a consequence of backscattering at either the contacts with leads or 
magnetic/potential barriers in the wire.  
\end{abstract}


\maketitle

\section{\label{sec:intro} Introduction}

Design and technological realization of quantum nanodevices requires
nanoscale systems of well defined and controllable properties. Recently,
tubular semiconductor structures turned out to be promising building
blocks of such appliances. Nanotubes of very narrow, but finite thickness
may be achieved in few different ways. In the case of quantum wires built of
narrow-gap materials surface states may induce Fermi level pinning above
the conduction band edge which results in accumulation of electrons in the
vicinity of the surface \cite{Luth10}. Nowadays it has become feasible
to combine two (or even more) different materials into one vertical
structure, i.e., core-shell nanowires (CSNs). This provides a possibility
to achieve thin tubular shells surrounding a core nanowire or other shells. 
One of the advantages of these systems is a possibility to establish band
alignment through the thicknesses of the components \cite{Shi15,Pistol08,Wong11}
and thus grow structures in which electrons are confined only in narrow
shell areas \cite{Jacopin12,Blomers13}. Moreover, the core part may be
etched such that separated nanotubes are formed \cite{Rieger12, Haas13}.

Most commonly CSNs have hexagonal cross-sections
\cite{Blomers13,Rieger12,Haas13}, but triangular \cite{Qian04,Qian05}
and circular \cite{Richter08} systems have also been achieved.  Electrons
confined in prismatic CSNs may form conductive channels along the
sharp edges \cite{Jadczak14,Bertoni11,Royo14,Royo15,Fickenscher13,Shi15,Sitek15}.
Rich quantum transport phenomena have been observed in CSNs,
e.g., flux periodic, similar to Aharonov-Bohm (AB), magnetoconductance oscillations
\cite{Blomers13,Gul14}, single electron tunneling, or electron
interference \cite{Richter08,Blomers13}. Very interesting effects have
been predicted in the presence of a strong magnetic field perpendicular
to the wire axis. In particular, the field induces a 
complex topology of the electronic states.
Low-energy electrons may be found around two channels
along the CSN axis where the radial component of the field vanishes by
changing sign. Carriers on both sides of the lines are deflected towards
opposite directions and thus confined into so called snaking states
\cite{Ferrari08,Ferrari09,Royo13,Manolescu13}.  Higher energy electrons
start to occupy Landau states and form cyclotron orbits localized in
the areas where the radial component of the field takes maximal values.
With increasing energy electrons move towards the sample ends and form
edge states \cite{Manolescu13}.  To the best of our knowledge experimental
investigation of magnetotransport in ballistic CSNs with magnetic field transversal
to the nanowire, and of the effects of snaking states, have only recently
been attempted \cite{Heedt14}.

In this paper we focus on thin cylindrical conductive shells since in such
systems carrier localization or conductive channels are induced only by
an external magnetic field and thus such samples allow to observe purely
magnetic effects.  According to our recent calculations the existence of
snaking states leads to resolved resonances of the conductance when the
CSN is weakly coupled to external leads \cite{Rosdahl15}. In the present
paper we extend these results and analyze signatures of snaking states
in the conductance for wide range of sample-lead coupling strength which
can be controlled by variable potential barriers. We show that 
interference of the snaking states, due to backscattering from magnetic
or potential barriers, may lead to flux periodic magnetoconductance
oscillations, detectable in transport experiments.

The paper is organized as follows. We define the model of the system in 
Sec.\ \ref{sec:model} and describe the computational method in 
Sec.\ \ref{sec:method}. The results and their discussion are presented 
in Sec.\ \ref{sec:cond} while the final remarks are contained 
in Sec.\ \ref{sec:conclusions}.

\section{\label{sec:model} The Model}

\begin{figure} 
\begin{center}
\includegraphics [width=8 cm] {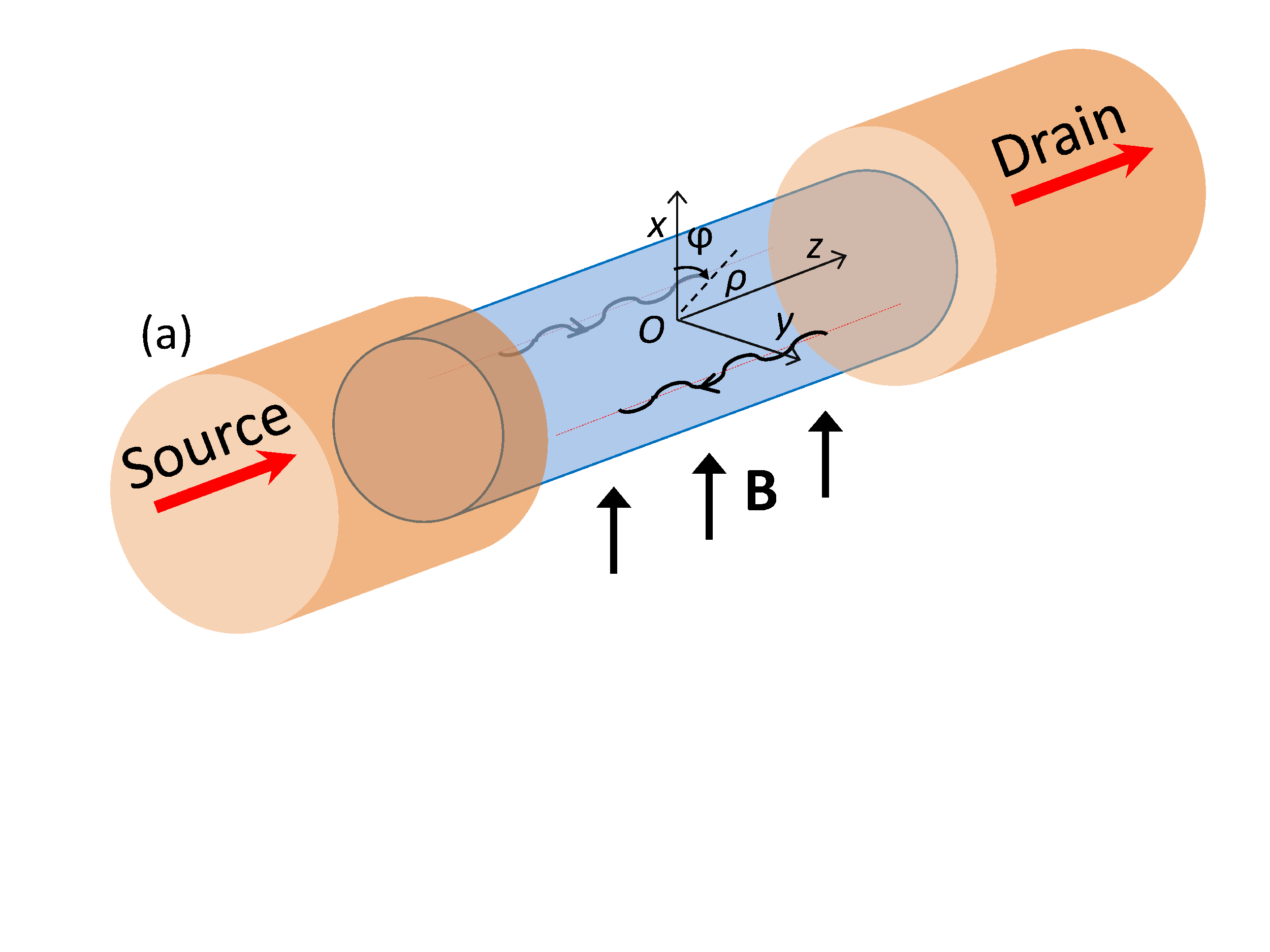}
\end{center}
\vspace{-30mm}
\begin{center}
\hspace{20mm}\includegraphics [width=5 cm] {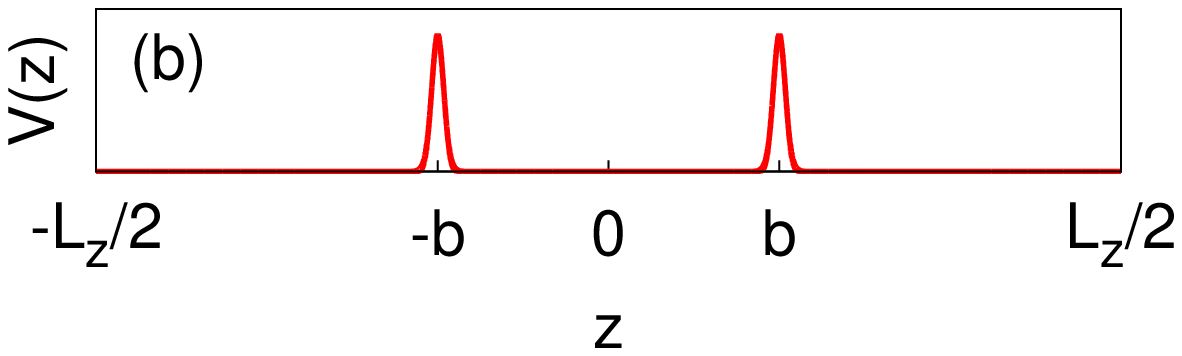}
\end{center}
\vspace{-5mm}
\caption{(Color online) (a) The CSN is a cylindrical surface of radius
$\rho$ and length $L_z$ nm (blue color).  The wavy lines with
arrows indicate snaking orbits propagating along the axes of zero radial
magnetic field (thin red dotted lines).  The contacts with the 
source and the drain electrodes are shown (gold color).
(b) Gaussian potential barriers of variable height 
placed along the CSN.  }
\label{Sample}
\end{figure}

The model of a CSN used here is a simple cylindrical surface of radius
$\rho$ and length $L_z$, through which a current can flow
from one end to another due to a potential bias.  We treat the electrons
like a cylindrical two-dimensional electron gas.  We consider a magnetic
field perpendicular to the axis of the cylinder.  When the field is
sufficiently strong low-energy electrons can travel along the snaking
orbits created along the two lines of zero radial component, at
polar angles $\varphi=\pi/2$ and $\varphi=3\pi/2$, as illustrated
in Fig.\ \ref {Sample}(a).  Two leads are attached to the CSN, one at
each end of it, which are treated as particle reservoirs, without a
specified shape.  Conventionally, we call them  Source (S) and Drain (D).


Two potential barriers are placed along the nanowire, at $z = \pm b$, 
symmetrically around the center $z = 0$.  The barriers are defined
as Gaussian functions $V(z)=V_0\exp{[-((z-b)/c)^2]}$ with a width
parameter $c$, and height $V_0$, Fig.\ \ref{Sample}(b).  The potential
barriers are independent on the polar angle $\varphi$.  Their role is 
to either control the contact strength, if placed at the contacts 
between the nanowire and the leads, and/or to contribute to the 
backscattering of the wave functions.


%

\section{\label{sec:method} The Computational Method}

In order to calculate the conductance of the open cylinder, i.e. the CSN
in contact with the external leads, we use the scattering formalism based
on the R-matrix method. This method has been used in similar transport
problems in quantum dots connected to the leads via quantum point contacts
\cite{Racec10}, in planar nanotransistors \cite{Nemnes04,Nemnes05}, 
or in cylindrical bulk nanowires with radial constrictions,
at zero magnetic field \cite{Racec09,Racec14,Nemnes09,Nemnes10}.
The approach consists of two parts.  In the first part the wave function
of an electron at a given energy $E$ is built, both in the leads and in
the central scattering region (here the CSN), and matched by continuity
conditions at the junctions.  In the second part one obtains the $S$
matrix, the transmission function, and the conductance.

We assume that in the leads, close to the junctions with the CSN, the wave
function can be written as a combination of plane waves and orbital states
\begin{equation}
\psi_l({\bf r}) \! = \! 
\sum_{m\sigma}\! \left[ \psi_{m\sigma l}^{in}e^{-ik_m(z-z_l)} 
+ \psi_{m\sigma l}^{out}e^{ik_m(z-z_l)} \right] \! u_m(\varphi) |\sigma\rangle, 
\label{psi_l} 
\end{equation}
where $l=S,D$ is a label for the two leads, $z_l=\mp L_z/2$ denote
the coordinates of the junctions, $\sigma$ is the spin label,
$u_m(\varphi)=e^{im\varphi}/\sqrt{2\pi}$ are the eigenvectors of the
angular momentum, with $m=0,\pm 1,\pm 2, ...$. The wavevector $k_m$ 
corresponds to the longitudinal motion of a 
particular circular mode $m$, being determined by the energy of the
incoming electron,
\begin{equation}
E=\frac{\hbar^2 }{2 m_{\rm eff}}\left(k_m^2 + \frac{m^2}{\rho^2}\right), \,\,\,\,
{k_m =\sqrt{\frac{2 E m_{\rm eff}}{\hbar^2}-\frac{m^2}{\rho^2}}}\ ,
\label{km}
\end{equation}
%
%
where $m_{\rm eff}$ is the effective mass of the material.  
Note that at a fixed energy, depending on $m$, $k_m$ can be real or imaginary.
The real values describe open channels, propagating from one lead to 
another lead, whereas the imaginary values describe closed (or evanescent) 
channels which are states bound around the scattering region \cite{Sitek15}.  

Although we are formally treating the leads as semi-infinite extensions of
the CSN, with the same circular symmetry, in fact the wave functions (\ref{psi_l})
are important only at (or close to) the boundaries $z_l$.  Therefore,
in principle, the shape of the leads can be arbitrary.  In experimental
setups the leads are usually perpendicular "finger" electrodes attached
to the nanowire sample \cite{Gul14}.  Moreover the magnetic field has
completely different effects in the leads and in the measured sample.
This also motivates us to neglect in our model the magnetic field in the
leads.  In addition to simplicity, this assumption is helpful to define
the contact between the leads and the CSN not only as a mathematical
boundary, but also as a magnetic barrier.  Then, by using the potential
barriers well inside the CSN we can also simulate new boundaries of the
scattering area, this time with a continuous magnetic field.

In order to calculate the wave function in the CSN region one has
to find the eigenstates of the Wigner-Eisenbud (WE) Hamiltonian $\tilde H$, satisfying
Neumann boundary conditions at the points $z_l$ (instead of the Dirichlet
conditions familiar for hard wall boundaries),
\begin{equation}
\tilde H\chi_a =  \epsilon_a \chi_a \ ,
\nonumber
\end{equation}
where $a=1,2,...$ is a generic quantum number labeling the WE energies
$\epsilon_a$ in increasing order. The WE eigenstates $\chi_a\equiv
|a\rangle$ are expanded in a basis set $|q\rangle=u_m(\varphi)u_n(z)|\sigma\rangle$,
with $u_n(z)=A_n\cos[n\pi(z/L_z+1/2)]$, $n=0,1,2,...$ and normalization
factor $A_0=\sqrt{1/L_z}$ and $A_n=\sqrt{2/L_z}$ for $n>0$.

The Hamiltonian $\tilde H$ is formally built as a regular
Hamiltonian $H$, with the kinetic term containing the modified momentum
operator $p_z+eA_z=p_z+eB\rho\sin\varphi$, the vector potential being defined in
the Landau gauge ${\bf A}=(0,0,By)$, and
the Zeeman term depending on the effective g-factor of the material,
\begin{equation}
H =  -\frac{\hbar^2}{2m_{\rm eff} \rho^2}\frac{\partial^2}{\partial\varphi^2}+
\frac{(p_z+eB\rho\sin\varphi)^2}{2m_{\rm eff}}-\frac{1}{2}g_{\rm eff}\mu_BB\sigma_x \ .
\nonumber
\end{equation}
With Neumann boundary conditions, implemented via the cosine functions
of the basis $|q\rangle$, the resulting linear term in $p_z$ is
not Hermitean.  Therefore the matrix elements of the WE Hamiltonian
are defined as $\tilde H_{qq'}=\left( H_{qq'} + H_{q'q}^* \right)/2$.
This procedure is equivalent to correcting the momentum $p_z$ with a
surface Bloch operator $L=-i\hbar[\delta(z-z_D)-\delta(z-z_S)]/2$ as
discussed by other authors \cite{Jayasekera06,Varga09}.

The wave function in the CSN can be written as a superposition of 
WE eigenstates, 
\begin{equation}
\psi({\bf r}, E)=\sum_a \alpha_a(E)\chi_a{(\bf r)},
\label{WFSR}
\end{equation}
and the coefficients $\alpha_a$ are determined by the continuity conditions
of the wave functions and their first derivatives at the $z_l$ boundaries.  Detailed
calculations can be found in Appendix A of Ref.\ \cite{Nemnes04}.
By introducing one more composite label $|\nu\rangle=|m\sigma l\rangle$ the amplitudes
of the wave function in the leads, Eq.\ (\ref{psi_l}), can be seen as the vectors
$\psi^{in}\equiv\{ \psi_{\nu}^{in} \}$ and 
$\psi^{out}\equiv\{ \psi_{\nu}^{out} \}$, which are related via
the continuity conditions, as
\begin{equation}
\psi^{in} + \psi^{out} = -i R K\left( \psi^{in} - \psi^{out}\right) \ .  \\ 
\label{conteq}
\end{equation}
In Eq.\ (\ref{conteq}) we have introduced two matrices, the matrix of wavevectors 
with elements
$K_{\nu\nu'}=k_m\delta_{\nu\nu'}$, and the so-called $R$ matrix defined as 
\begin{equation}
R_{\nu\nu'}(E)=-\frac{\hbar^2}{2m_{\rm eff}}\sum_a 
\frac{\langle \nu | a\rangle \langle {\nu' | a\rangle}^{\dagger}} {E-\epsilon_a} \ .
\label{Rmat}
\end{equation}
The notation $\langle \nu | a\rangle$ stands for the scalar product of the orbital 
and spin states incorporated in each factor, at the two frontiers $z_l$, i.e. 
\begin{equation}
\langle \nu | a\rangle = \langle\sigma| \int_0^{2\pi} u_m^*(\varphi) \chi_a(\varphi,z_l) d \varphi \ .
\nonumber
\end{equation}

The scattering problem is solved by calculating the $S$ matrix, which transforms 
the "in" states in "out" states, 
\begin{equation}
\psi^{out} = S \psi^{in} \, ,\\
\nonumber
\end{equation}
and using Eq.\ (\ref{conteq}) it is obtained as
\begin{equation}
S = - (1-iRK)^{-1}(1+iRK) \ .
\nonumber
\end{equation}
Having the $S$ matrix one can calculate the transmission matrix between 
the open channels $\nu$ and $\nu'$ 
\begin{equation}
T_{\nu\nu'}(E)=  \left| \left( K^{1/2} S K^{-1/2} \right )_{\nu\nu'} \right|^2 \ ,
\nonumber
\end{equation}
and finally the conductance $G$, by summing all contributions
from separate leads, i.e.  $|\nu\rangle=|m\sigma S\rangle$ and
$|\nu'\rangle=|m'\sigma' D\rangle$,
\begin{equation}
G=\frac{e^2}{h}\int dE \left( -\frac{\partial \cal F}{\partial E} \right) 
\sum_{\substack {m \sigma \\ m \sigma'}} T_{m\sigma S,m'\sigma' D}(E) \ ,
\label{cond}
\end{equation}
where ${\cal F}$ denotes the Fermi function.

\section{\label{sec:cond} Results and discussion}

In the numerical calculations we used material parameters of InAs,
$m_{\rm eff}=0.023$ and $g_{\rm eff}=-14.9$.  The radius of the CSN was
fixed to $\rho=30$ nm and length was $L_z=300$ or $2000$ nm.  The results
were convergent in a basis $|q\rangle$ truncated to orbital momenta with
$|m| \leq 10$ and longitudinal modes $n\leq 130$.  All channels, both
open and closed, were used to calculate the $R$ and $S$ matrices. The
temperature was fixed to $T=0.5$ K.

\subsection{\label{sec:zeromag} Zero magnetic field}

\begin{figure} 
\begin{center}
\includegraphics [width=86 mm] {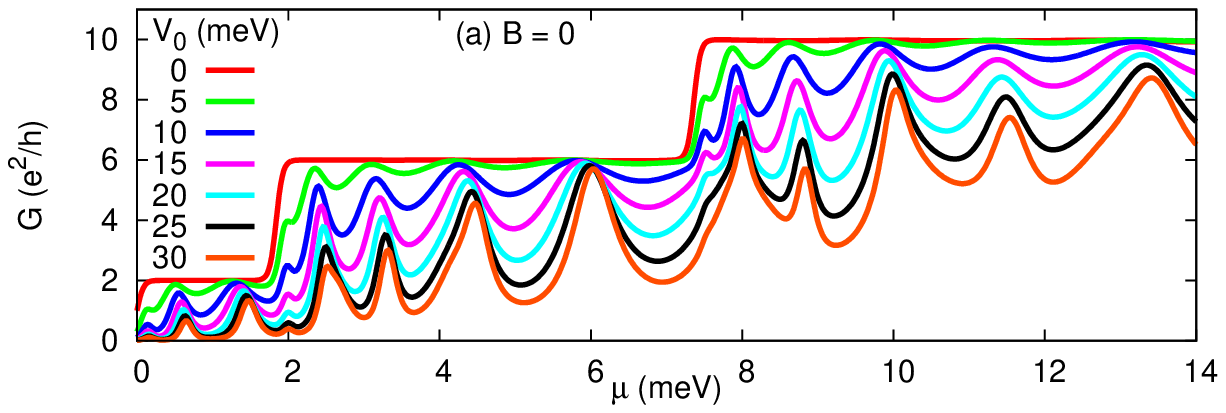}
\includegraphics [width=86 mm] {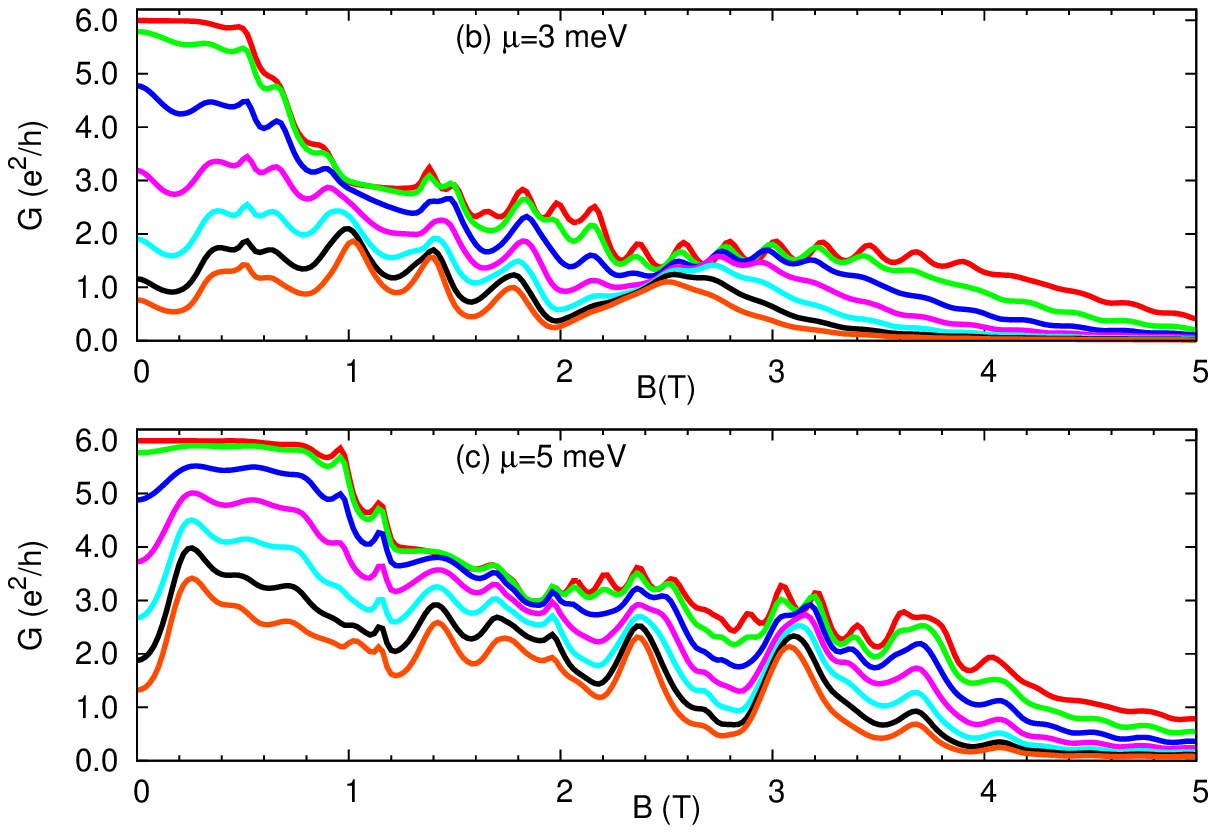}
\end{center}
\vspace{-5mm}
\caption{(Color online) 
(a) Conductance vs. chemical potential $\mu$ without magnetic field,
for contact barriers of height $V_0=0,5,...,30$ meV,
and then vs. magnetic field perpendicular to the nanowire, with
(b) $\mu=3$ meV and (c) $\mu=5$ meV.
The nanowire length is $L_z=300$ nm and the barriers are close to 
the contacts, at $b=\pm 147.5$ nm, of width $c=2.5$ nm.
}
\label{Bx_var}
\end{figure}

First we show in Fig.\ \ref{Bx_var}(a) the conductance at zero magnetic
field as a function of the chemical potential $\mu$ for several
heights of the potential barriers. In this case $L_z=300$ nm,
and the barriers are situated very close to the contacts, 
at $b=\pm 147.5$ nm, having a width of $c=2.5$ nm. 
The results are as expected for
ballistic transport in a quantum wire.  Without the barriers ($V_0=0$)
the conductance has the familiar steps given by the number of open
channels, which is the number of $m$ values yielding a real wavevector $k_m$
for the energy $E=\mu$ in Eq.\ (\ref{km}), multiplied by the two spin
states $\sigma=\pm 1$.  By increasing the height of the barriers the
conductance drops, because the energy of the electronic states within 
the CSN increases, and thus a smaller
number of open channels remains available up to the fixed Fermi level. Also
the conductance begins to oscillate, as a results of the (Fabry-Perot) interference
between transmitted and reflected waves, and evolve towards resonances
with peaks indicating the density of states in the scattering region.
This is a consequence of the fact that the coupling between the scattering
region and the leads decreases.
%

\subsection{\label{sec:magbar} Magnetic barriers with adjustable contacts}

Another way to create backscattering of the electrons at contacts is to
consider a magnetic barrier, i.e. a magnetic field that exists only in
the CSN, but not in the leads, as we assumed in Section \ref{sec:method}.  In 
addition, by adding the potential barriers we can modify the contact strength.
In Fig.\ \ref{Bx_var}(b), we show how the conductance depends
on the strengths of the transverse magnetic field and on the coupling to
the leads for chemical potential 3 meV.  We distinguish two
regimes, corresponding to low and high potential barriers, respectively.
For weak potentials, i.e. $V_0$ up to 5 meV, regular conductance
oscillations are obtained, with periods $\Delta B$ slightly increasing
from 0.20 to 0.22 T in the interval $B=2\textendash 4$ T.  Each of
these oscillations nearly corresponds to a gain of one flux unit
$\Phi_0=h/e$ through the area of the cylinder projected on the $yz$
plane, $A=2\rho L_z$. According to this estimation the period should
be $\Delta B=\Phi_0/A=0.23$ T.  Therefore these oscillations can be
considered a kind of AB interference of snaking states
propagating on the lateral sides of the cylinder.

By increasing the height of the potential barriers the AB-like oscillations
smear out and the broader conductance peaks emerge.  These peaks are
produced by the same snaking states, but now as individual resonances
occurring in the nearly isolated CSN, only weakly connected to the
leads. This case was described by Rosdahl et al. by modeling the contacts
with a tunneling parameter \cite{Rosdahl15}.

In Fig.\ \ref{Bx_var}(c) we show the magnetoconductance with the chemical
potential increased to 5 meV.  The contribution of higher energy levels
results in more complex fluctuations, but the fine structure of the flux periodic 
oscillations remains visible.  The resonances at high potential barriers
are now shifted to higher magnetic fields.

\subsection{\label{sec:spewf} WE Energy spectra and wave functions}

\begin{figure} 
\begin{center}
\includegraphics [width=86 mm] {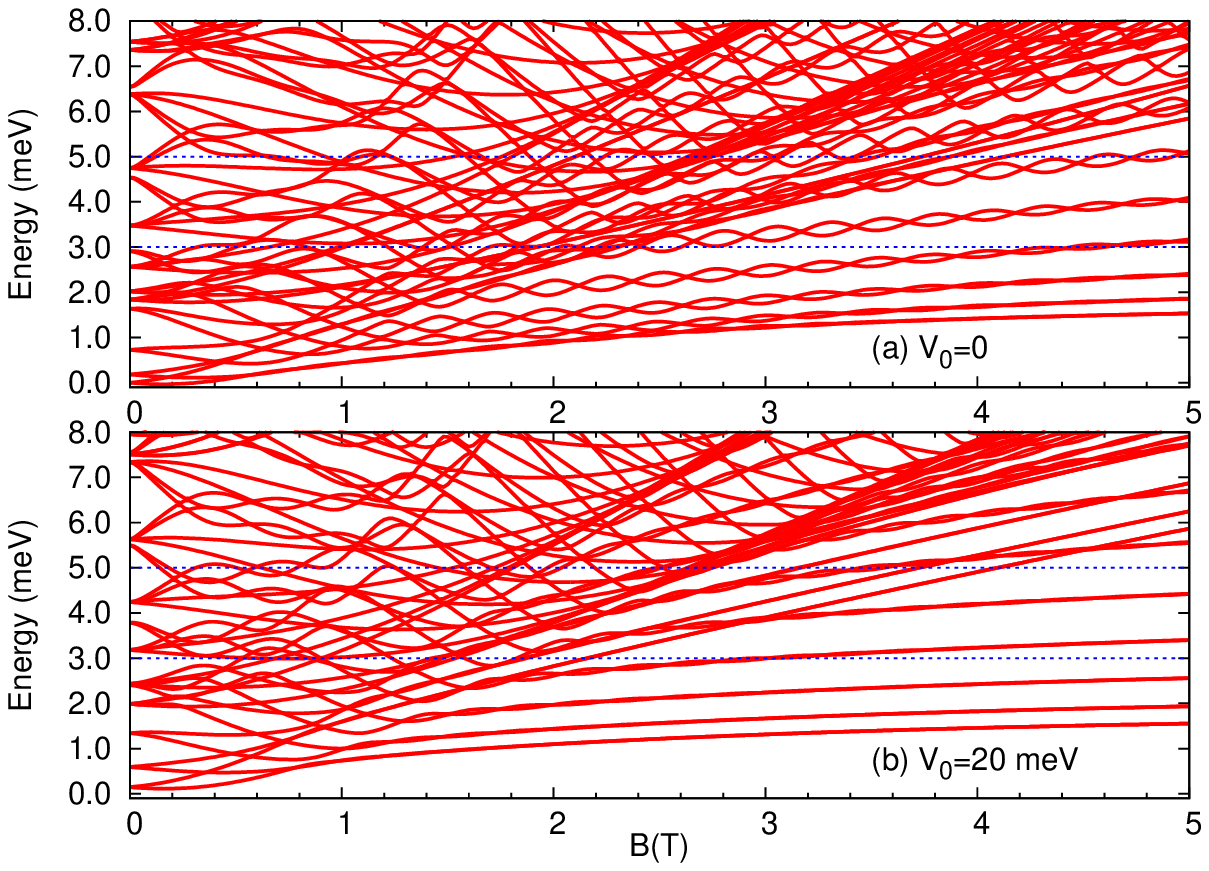}
\end{center}
\vspace{-5mm}
\caption{(Color online) The Wigner-Eisenbud (WE) energies $\epsilon_a$ 
(a) without contact barriers and
(b) with barriers of $V_0=20$ meV.  The blue dotted horizontal lines
show the chemical potentials used in the magnetoconductance calculations,
$\mu=3$ meV and $\mu=5$ meV.  }
\label{WEspectra}
%
%
\begin{center}
\vspace{-8mm}
\includegraphics [width=88 mm] {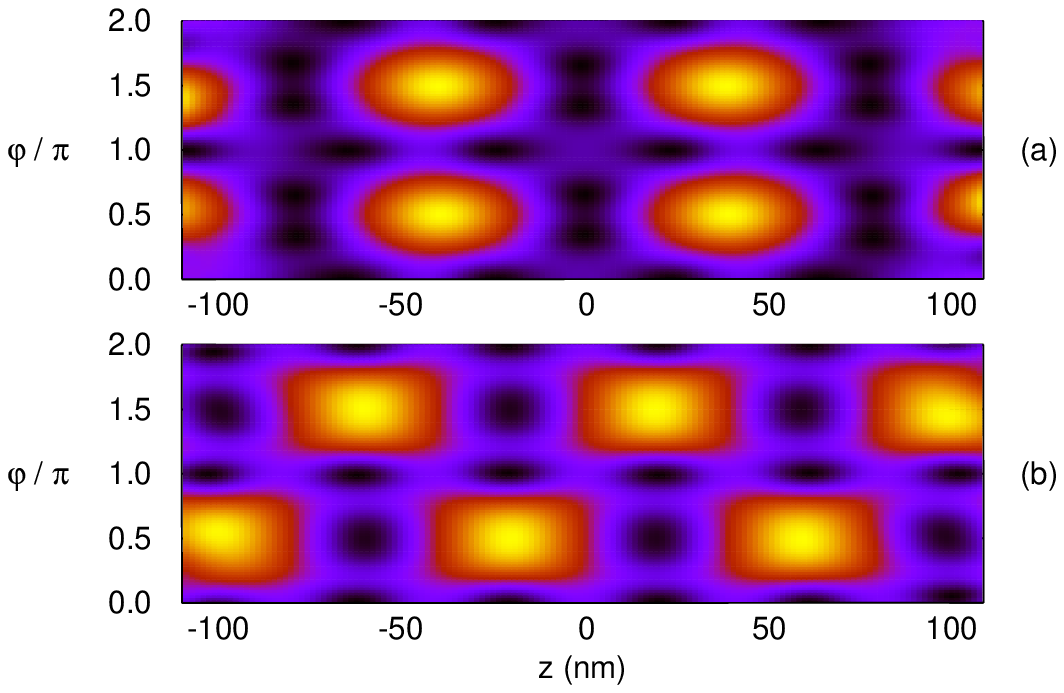}
\end{center}
\vspace{-15 mm}
\caption{(Color online) Probability densities corresponding to the wave functions
within the CSN, at energy $E=3$
meV and $V_0=0$, for (a) $B=2.91$ T when the snaking
states are in phase, and (b) $B=3.02$ T when they are in anti-phase.
Here we show the open cylindrical surface,
the vertical axis being the polar angle and the horizontal axis
the longitudinal $z$ coordinate. }
\label{WF}
\end{figure}

We can gain more understanding of the AB oscillations by looking at the
WE energy spectra vs magnetic field shown in Fig.\ \ref{WEspectra}.
In the case of strong coupling between the CSN and the leads, e.g. 
$V_0=0$, the low energy levels form braid shape patterns for $B>1$ T,
Fig.\ \ref{WEspectra}(a).  These oscillations affect the conductance,
Eq.\ (\ref{cond}), through the denominators of the the R-matrix, Eq.\
(\ref{Rmat}), which are sensitive to such small changes in the WE energies
$\epsilon_a$, and induce the AB oscillations.  In fact, the R-matrix has
a form similar to the Green functions used by other authors for such
scattering-transport calculations \cite{Tserkovnyak06}.  The braids are
an indication of snaking states interference in the open CSN.  We verified
that such an energy spectrum is also obtained for a CSN with a finite 
thickness of 10 nm, by including in the basis $|q\rangle$
radial wave functions vanishing at the surfaces.

In the presence of high potential barriers the braids shrink and converge
towards nearly double degenerate eigenstates, Fig.\ \ref{WEspectra}(b),
as obtained for the isolated cylinder \cite{Rosdahl15}.  The WE spectra
can also explain the transition from flux-periodic magnetoconductance
to resonant peaks, observed while the height of the potential barriers is
being increased. These peaks occur at those magnetic fields for which
the snaking states are crossing the Fermi energy.
%

Further information on the snaking states can be obtained from the wave functions.
In the scattering region they are obtained with Eq.\ (\ref{WFSR}), using
the coefficients $\alpha_a$ expressed with the $S$ matrix \cite{Racec09},
\begin{equation}
\alpha_a(E) \! = \! \frac{\hbar^2}{2m_{\rm eff}}\frac{i}{\sqrt{2\pi}}
\frac{1}{E-\epsilon_a} \!
\sum_{\nu}{\langle\nu|a\rangle}^*
k_m \!\! \left( \! 1- \! \sum_{\nu'}S_{\nu\nu'} \!\! \right),
\nonumber
\end{equation}
where by summing over all labels $\nu$ we consider electrons incoming
from both leads, with wave functions of equal amplitudes. 
Nevertheless, since the wave functions themselves are not directly
involved in the conductance calculations, we look at them only to correlate
the behavior of the snaking states in the scattering region 
with the WE spectrum and the conductance oscillations. 

In Fig.\ \ref{WF} we show two distinct situations, for a fixed
energy $E=3$ meV, which is the chemical potential used in Fig.\
\ref{Bx_var}(b), and no potential barrier, $V_0=0$.  For $B=2.91$ T the
pair of snaking states have in-phase longitudinal oscillations, Fig.\
\ref{WF}(a), corresponding to a crossing of the braided WE energies, Fig.\
\ref{WEspectra}(a), and to a conductance minimum, Fig.\ \ref{Bx_var}(b).
For $B=3.02$ T the two snaking states have out-of-phase longitudinal
oscillations, Fig.\ \ref{WF}(b). In this case the braids are maximally
open, and the conductance has a maximum.  The out-of-phase structure
of the snaking states does not exist in the isolated CSN, but only the
in-phase one \cite{Rosdahl15}.  Here, for the open CSN, by imposing the
potential barriers the lateral shift of the out-of-phase maxima gradually 
reduces, until they align like in Fig.\ \ref{WF}(a).  Seen from this 
angle the braids of the WE spectrum and the AB oscillations are related
to the relative phase of the snaking states.

\subsection{ Potential barriers on a long wire} 

\begin{figure} 
\begin{center}
\includegraphics [width=86 mm] {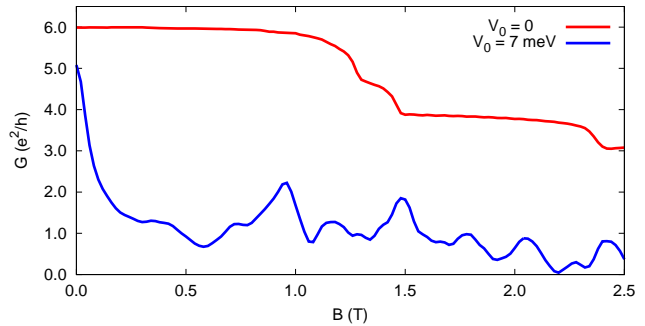}
\end{center}
\vspace{-5mm}
\caption{(Color online) Conductance vs. magnetic field 
for a nanowire of length $L_z=2000$ nm. The upper curve was 
obtained without potential barriers ($V_0=0$), and the lower curve
with two barriers situated at $b=\pm 150$ nm, as indicated in 
Fig. \ref{Sample}(b), of height $V_0=7$ meV, and width parameter $c=20$ nm.
The chemical potential is $\mu=6.5$ meV.}
\label{Long_wire}
\end{figure}

In the next example we chose a much longer CSN, of length $L_z=2000$ nm.
The magnetic field still vanishes in the leads, and so there are
magnetic barriers at the contacts, which cause backscattering and hence
interference. However, the corresponding conductance oscillations are now too
dense and too weak to be resolved, because their period is inversely
proportional to the wire length.  In the absence of
any potential barrier the computed magnetoconductance has smooth steps
as shown in Fig.\ \ref{Long_wire}.  The steps reflect the number of 
propagating (open) channels associated to the complex subband structure
of the energy spectra of an infinite hollow cylinder in transverse magnetic 
field \cite{Ferrari08,Ferrari09,Manolescu13,Tserkovnyak06}.  A detailed
analysis of these steps is nevertheless beyond the aim of the present
paper. 

With potential barriers situated at $\pm 150$ nm from the center of the
nanowire we can obtain again the flux periodic oscillations, comparable
to the ones for the nanowire of 300 nm length.  The oscillations are now
less regular than before, but with an average period of 0.20-0.22 T within
the interval 0.4-2.4 T.  In order to reduce the transparency of the barriers 
and to increase the backscattering we used wider barriers than before,
with $c=20$ nm.  The length of the scattering zone in between the
potential barriers is less sharply defined than in the case of the pure magnetic
barriers, and so is the real area of the magnetic flux, compared to the
reference cross section area between the barriers, $A=4\rho b$. Nevertheless, 
regardless of these imperfections, the oscillations correspond reasonably
well with the expected periodicity $\Delta B = \Phi_0/A=0.23$ T. 

\vspace{10mm}

\section{\label{sec:conclusions} Conclusions}

In conclusion we predict that the existence of the snaking states in a
CSN in a transversal magnetic field can be experimentally observed as
flux periodic oscillations of the magnetoconductance, with a short CSN 
strongly coupled to leads.  In this case the snaking
states behave like transmitted and reflected waves which interfere at
the contacts with the leads.  In the limit of weak coupling the snaking
states can be seen as individual resonances of the conductance.  In our
model the contacts are primarily simulated by matching two different types
of wave functions, in the presence of a magnetic barrier resulting from
neglecting the magnetic field in the leads.  In order to further modify
the transmission and reflection at the contacts we included potential
barriers, which reduced the coupling CSN-leads, and thus the amplitude
of the transmitted waves.  Another way to observe the flux periodic
oscillations in a transverse magnetic field, although possibly less
regular, may be by creating scattering regions with potential barriers,
as produced by finger gates placed over a long nanowire \cite{Gul14}.


\begin{acknowledgments}
This work was financially supported by the research funds of Reykjavik
University and of the University of Iceland, and by the Icelandic
Research Fund. T. O. Rosdahl acknowledges support from an ERC
Starting Grant. We are thankful to Thomas Sch\"apers and Sebastian
Heedt for very interesting discussions \cite{Heedt14}.
\end{acknowledgments}

\bibliographystyle{apsrev4-1}
%

\end{document}